\documentclass[twocolumn]{aastex631}
\usepackage{amsmath}
\usepackage{array}
\usepackage{graphicx}
\usepackage{algorithm}
\usepackage{algorithmic}

\shorttitle{Implications of a Possible Spectral Structure of Cosmic-Ray Protons }
\shortauthors{Nie et al.}
\begin{document}
%\linenumbers

\title{Implications of a Possible Spectral Structure of Cosmic-Ray Protons Unveiled by the DAMPE}

\correspondingauthor{Ze-Jun Jiang}
\email{zjjiang@ynu.edu.cn}
\author{Lin Nie}
\email{nielin@mail.ynu.edu.cn}
\affiliation{Department of Astronomy, Yunnan University, and Key Laboratory of Astroparticle Physics of Yunnan Province, Kunming, 650091, People’s Republic of China.}
\author{Yang Liu}
\affiliation{Department of Astronomy, Yunnan University, and Key Laboratory of Astroparticle Physics of Yunnan Province, Kunming, 650091, People’s Republic of China.}
\author{Zejun Jiang}
%\email{zjjiang@ynu.edu.cn}
\affiliation{Department of Astronomy, Yunnan University, and Key Laboratory of Astroparticle Physics of Yunnan Province, Kunming, 650091, People’s Republic of China.}

\begin{abstract}
 The recent observations revealed that the cosmic-ray (CR) proton spectrum showed a complex structure: the hardening at $\rm \sim 200\,GeV$ and softening at $\rm \sim 10\,TeV$. However, so far the physical origins of this spectral feature remain strongly debated. In this work, we simulate the acceleration of cosmic-ray protons in a nearby Supernova remnant (SNR) by solving numerically the hydrodynamic equations and the equation for the quasi-isotropic CR momentum distribution in the spherically symmetrical case to derive the spectrum of protons injected into the interstellar medium (ISM), and then simulate the propagation process of those accelerated CR particles to calculate the proton fluxes reaching the Earth. Besides, we use the DRAGON numerical code to calculate the large-scale cosmic-ray proton spectrum. Our simulated results are in good agreement with the observed data (including the observed data of proton fluxes and dipole anisotropy). We conclude that the spectral feature of cosmic-ray protons in this energy band may originate from the superposition of the distribution from the nearby SNR and background diffusive cosmic-ray component. We find that the release of particles from this nearby SNR has a time delay. Besides, it can be found that the nonlinear response of energetic particles, release time of CR protons, and age of the local SNR can leave strong signatures in the spectrum of the resulting CR proton fluxes.
%\textbf{}
\end{abstract}
%\keywords{Cosmic rays; Supernova remnant; Shocks}

\section{INTRODUCTION} \label{sec:introduction}
Supernova remnants (SNRs) are considered to be the most promising sites for the acceleration of Galactic cosmic rays since they can provide sufficient energy to maintain the cosmic ray energy fluxes in our Galaxy \citep{2020ApJ...894...51A}. Supernova remnants can in principle produce the source spectrum of Galactic cosmic rays required by the empirical model of cosmic ray origin \citep{2010ApJ...718...31P}. The clear evidence for cosmic-ray particle acceleration in SNRs has been given by observations of non-thermal radio, X-ray, and gamma-ray emissions \citep{2013Sci...339..807A}.

The observed spectrum of Galactic cosmic rays is believed to arise from  a combination of diffusive shock acceleration occurring in SNRs and diffusive transport off magnetic turbulence \citep{2007ARNPS..57..285S,2015ARA&A..53..199G,2014BrJPh..44..415B,2013A&ARv..21...70B}. After cosmic-ray particles are accelerated by the diffusive shock inside the SNRs, they are released by their sources. Then they enter the Milky way and interact with irregular magnetic fields and interstellar gas, which could be described as a diffusion process \citep{2021arXiv210700313L}.

The propagation of cosmic rays in the Milky way is a fundamental question in understanding the origin and interactions of Galactic CRs. The conventional cosmic-ray propagation model has some assumptions such as homogeneity, isotropy, and stationarity, and predicts that the observed energy spectrum falls as a featureless power law, e.g., $\propto R^{-\nu-\delta}$, with $\nu$ and $\delta$ being the power law indexes of injection spectrum and diffusion coefficient, respectively \citep{2015PhRvD..92h1301T}. However, more and more observations disfavor such a simple picture \citep{2023arXiv230506948Z,2022SciBu..67.2162D,2021PhR...894....1A,2018PhRvD..97f3008G,2012ApJ...752L..13T}. A series of new and more precise experiments, such as PAMELA, AMS-02, and DAMPE probing the fine structure of the CR spectrum, provide us with a useful tool to study the physical origins of different cosmic-ray spectral features \citep{2017PhRvD..95h3007Y,2020FrPhy..1624501Y}. This means that the measurements of the energy spectra of Galactic cosmic rays have entered a precise era. Several recent observations make the cosmic-ray proton spectrum show some new features: the hardening at $\rm \sim 200~ GeV$ and softening at $\sim 10\rm ~TeV$ \citep{2019SciA....5.9459A,2019SciA....5.3793A,2009BRASP..73..564P,2018JETPL.108....5A}. These observed phenomena generate a huge challenge to the standard model. The spatial-dependent propagation model (SDP) was first provided to describe the spectral hardening of primary cosmic-ray proton and helium above 200 GV which are observed by the AMS-02 experiments \citep{2020ChPhC..44h5102T}, but this would be still not sufficient to reproduce the softening feature at $E\rm \sim 10~TeV$. Therefore, many popular papers
start to consider a nearby source to explain both the hardening and softening spectral features of the cosmic-ray protons \citep{2021PhRvD.104j3013F,2020FrPhy..1624501Y}, in which the nearby source was assumed to have a power-law injection spectrum with a burst-like or continuous injection.
However, given that the complex plasma-flow profiles of SNR considerably modify the particle spectra, it is clear that this assumption cannot hold. This effect has been detected for a few SNRs \citep{2006ApJ...648L..33V}. In fact, the strong streaming of accelerated particles can change medium properties in the shock vicinity. When the particles are injected into shocks, the strong streaming results in significant shock modification, which will lead to a curvature of the particle spectrum with spectral hardening at the high energy\citep{2012APh....39...12Z}. At early epochs of the SNR expansion, the high injection efficiency can result in significant shock modification. With time, the Alfvenic heating upstream of the forward shock will result in a lower compression ratio and acceleration efficiency, so the shock modification is not strong\citep{2010ApJ...708..965Z}. Thus, the spectra of particles are steeper in comparison with ones at earlier epochs. Therefore, the SNR has a different injection spectrum at different epochs. To get reliable hydrodynamical data for the plasmas as well as good estimates for the cosmic-ray proton fluxes, self-consistently simulating the evolution of SNR and the acceleration process of CR particles inside SNR through solving the hydrodynamical equations to get the instantaneous injection spectrum of the nearby source is necessary.

In this work, we consider a nearby supernova remnant to provide an extra proton component to explain the spectral structure of cosmic-ray protons measured by the DAMPE. But different from the scenario described by the previous works \citep{2019JCAP...10..010L,2019JCAP...12..007Q,2021PhRvD.104j3013F}, to implement a more realistic injection spectrum of CR proton from the nearby SNR, we simulate the acceleration of cosmic-ray particles and nearby-SNR evolution. These cosmic-ray protons accelerated inside a nearby supernova escape from their source and inject into the interstellar medium and then diffuse to the Earth. Finally, we use the DRAGON numerical code to calculate the large-scale background proton spectrum \citep{2017JCAP...02..015E}.
\begin{figure}[t]
\centering
\includegraphics[width=\linewidth]{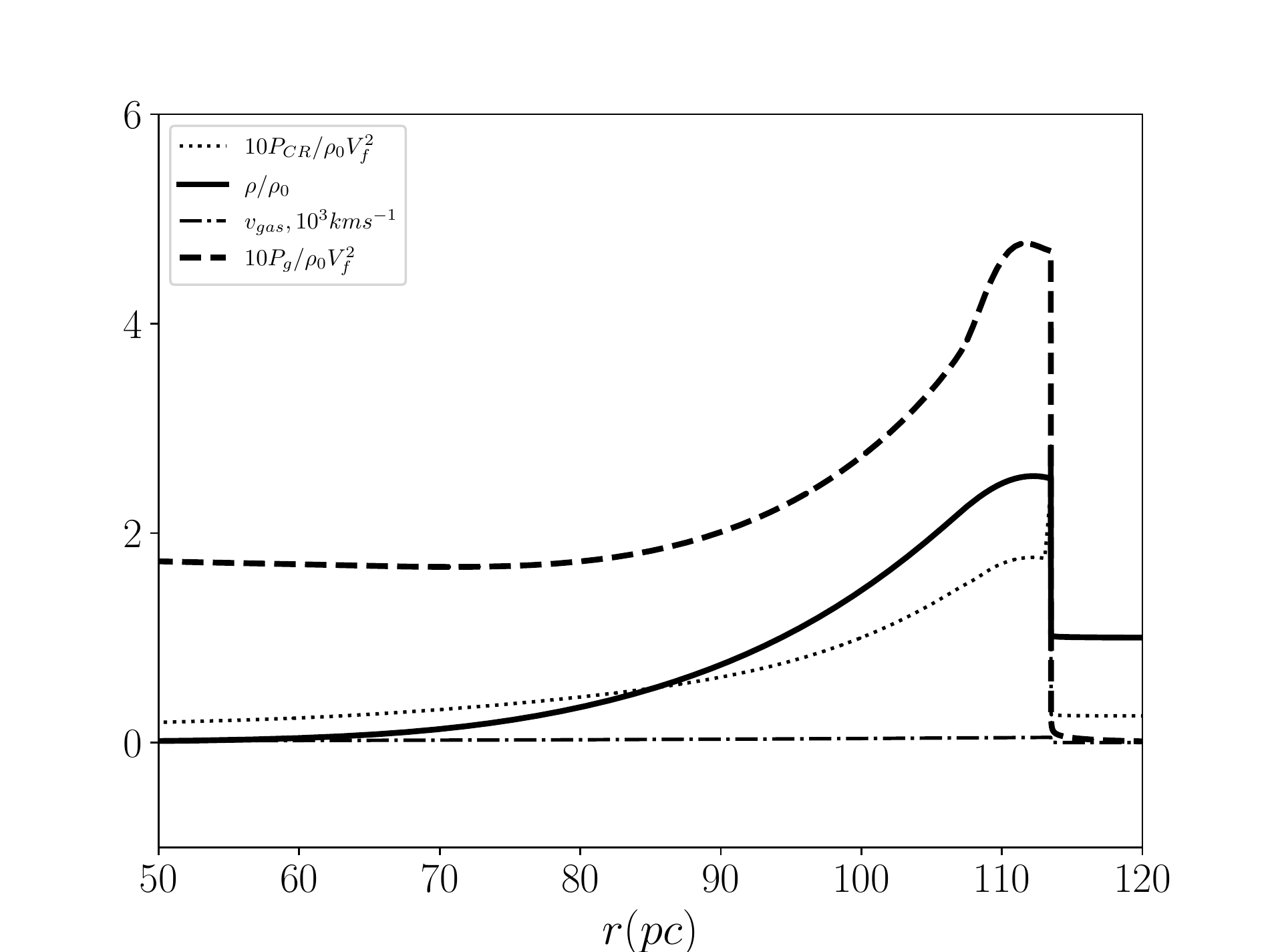}
\caption{The solid line represents radial dependencies of gas density; the dotted line shows cosmic ray pressure; the dashed-dotted line shows the gas velocity and the dashed line represents gas pressure at the SNR age of $t=7\times 10^5 \mathrm{yr}$.}
\label{figure1}
\end{figure}

\section{LARGE-SCALE CRs} \label{sec:background}
The Galactic CRs are accelerated inside cosmic-ray sources, such as supernova remnants \citep{2013Sci...339..807A,2014IJMPD..2330013A}, pulsar wind nebulae \citep{2021Natur.594...33C,2021Sci...373..425L,2022ApJ...924...42N}, and some other objects. After escaping into interstellar space, CRs diffuse within the Galaxy by randomly scattering off magnetic waves and magnetic hydrodynamic (MHD) turbulence. The corresponding propagation process could be described by a diffusion equation \citep{2017JCAP...02..015E,2007ARNPS..57..285S}
\begin{equation}
\begin{aligned}
\frac{\partial \psi}{\partial t} = & Q(\mathbf{x}, p)+\nabla \cdot\left(D_{x x} \nabla \psi-\mathbf{V}_{c} \psi\right)+\frac{\partial}{\partial p} p^{2} D_{p p} \frac{\partial}{\partial p} \frac{1}{p^{2}} \psi \\
&-\frac{\partial}{\partial p}\left[\dot{p} \psi-\frac{p}{3}\left(\nabla \cdot \mathbf{V}_{c} \psi\right)\right]-\frac{\psi}{\tau_{f}}-\frac{\psi}{\tau_{r}}
\end{aligned}
\end{equation}
where $\psi$ is the differential density of cosmic-ray particles per momentum interval, $Q$ is the source term, $D_{xx}$ is the spatial diffusion coefficient in the momentum space, $V_{c}$ is the convective velocity, $\tau_{f}$ and $\tau_{c}$ are correspondingly the time scales for fragmentation and radioactive decay.
\begin{figure}[t]
\centering
\includegraphics[width=\linewidth]{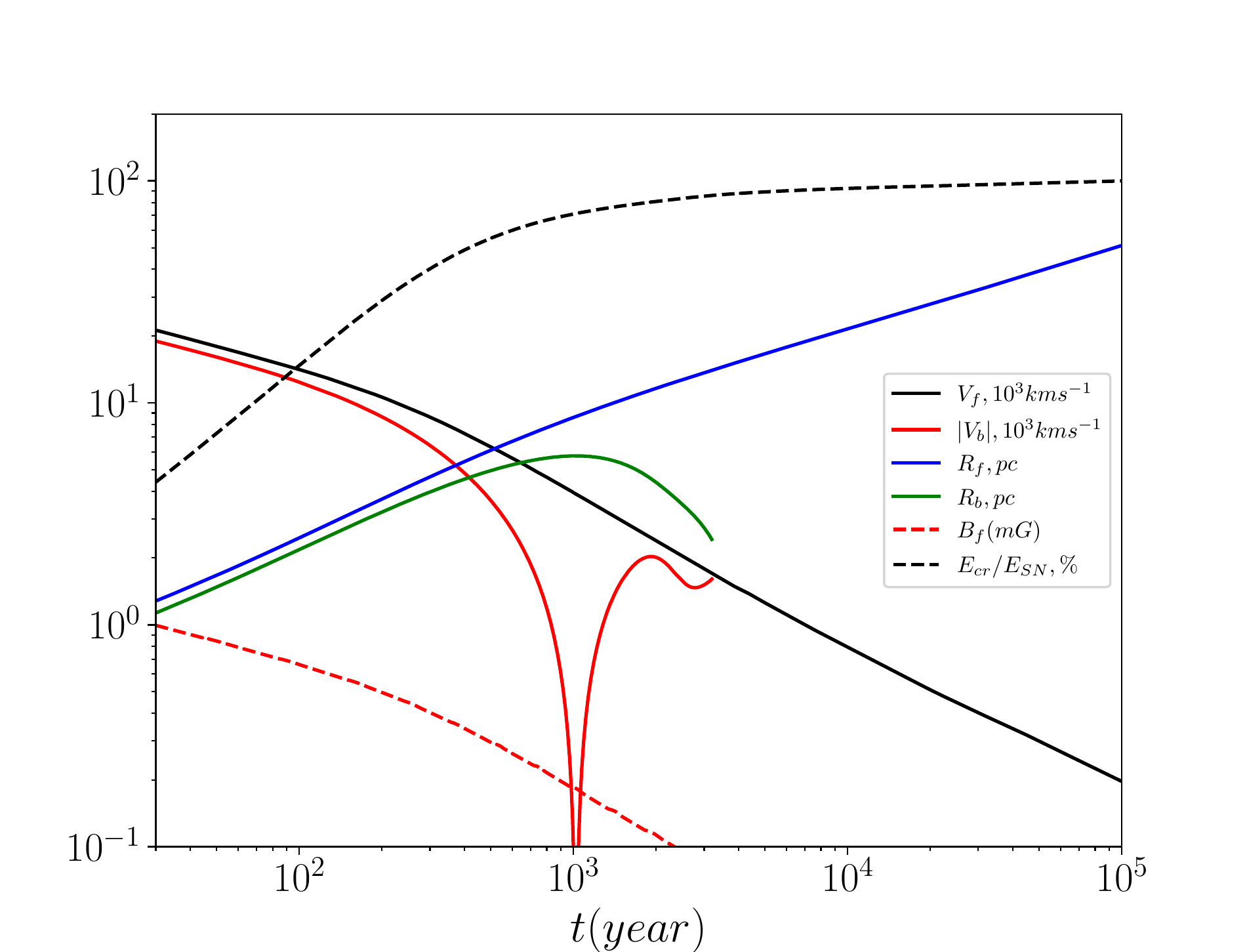}
\caption{The SNR forward shock (blue solid line) and reverse shock (green solid line) evolve with time; the black solid line and red solid line represent forward and reverse shock velocity, respectively. Finally, the magnetic field strength (red dashed line) downstream of the forward shock and the ratio of the CR energy and the total energy of supernova explosion (black dashed line) are also shown in the figure.}
\label{figure2}
\end{figure}

For the traditional diffusion scenario, the diffusion coefficient is assumed as uniform and isotropic cosmic-ray diffusion characterized by a spatial-independent scalar diffusion coefficient which follows a function of rigidity $D_{xx}(\rho)=D_{0}(\frac{\rho}{\rho_{0}})^{\delta_{0}}$. Where $D_{0}=D_{xx}(\rho_{0})$ is the diffusion coefficient normalization at a reference rigidity $\rho_{0}$. The spatial-dependent diffusion scenario is first described with a two-halo approach. It is believed that the level of turbulence is intense near the large population of sources due to the activities of supernova explosions \citep{2016ApJ...819...54G,2018PhRvD..97f3008G}. Therefore, the regions where the source density is high, the corresponding diffusion is slow. Hence the diffusion coefficient of the spatial dependent model is parameterized as a function of r and z,
\begin{equation}
D_{x x}(r, z, \mathcal{\rho})=D_0 F(r, z) \beta^\eta\left(\frac{\mathcal{\rho}}{\mathcal{\rho}_0}\right)^{\delta(r, z)}
\end{equation}
where $F(r, z)D_0$ is the normalization factor of the diffusion coefficient at the reference rigidity $\rho_0$, and $\delta(r, z)$ reflects the property of the irregular turbulence. They are parameterized as \citep{2018PhRvD..97f3008G,2018ApJ...869..176L}
\begin{equation}
F(r, z)= \begin{cases}g(r, z)+[1-g(r, z)]\left(\frac{z}{\xi z_h}\right)^n, & |z| \leqslant \xi z_h, \\ 1, & |z|>\xi z_h\end{cases}
\end{equation}
\begin{equation}
\delta(r, z)= \begin{cases}g(r, z)+\left[\delta_0-g(r, z)\right]\left(\frac{z}{\xi z_h}\right)^n, & |z| \leqslant \xi z_h \\ \delta_0, & |z|>\xi z_h\end{cases}
\end{equation}
Where $\xi z_h$ represents the half thickness of inner Galaxy halo. $g(r, z)=N_m /[1+f(r, z)]$, $N_m$ is a normalization factor, $n$ is used to characterize the sharpness between the inner and outer halos, and $f(r, z)$ is the source distribution. The spatial distribution of Galactic cosmic-ray sources are approximated as an axi-symmetric form, which can be parameterized as \citep{1996A&AS..120C.437C,1998ApJ...509..212S,1998ApJ...504..761C}
\begin{equation}
f(r, z)=\left(\frac{r}{R_{\odot}}\right)^{\alpha} \exp \left(- \frac{\beta(r-R_{\odot})}{R_{\odot}}-\frac{|z|}{z_{0}}\right)
\end{equation}
where $R_{\odot}=8.5 \rm~ kpc$ represents the distance from the Galactic center to the solar system. Parameters $z_0$, $\alpha$ and $\beta$ are fixed as $0.2 \rm~ kpc$, 1.69 and 3.33, respectively.

\begin{figure}[t]
\centering
\includegraphics[width=\linewidth]{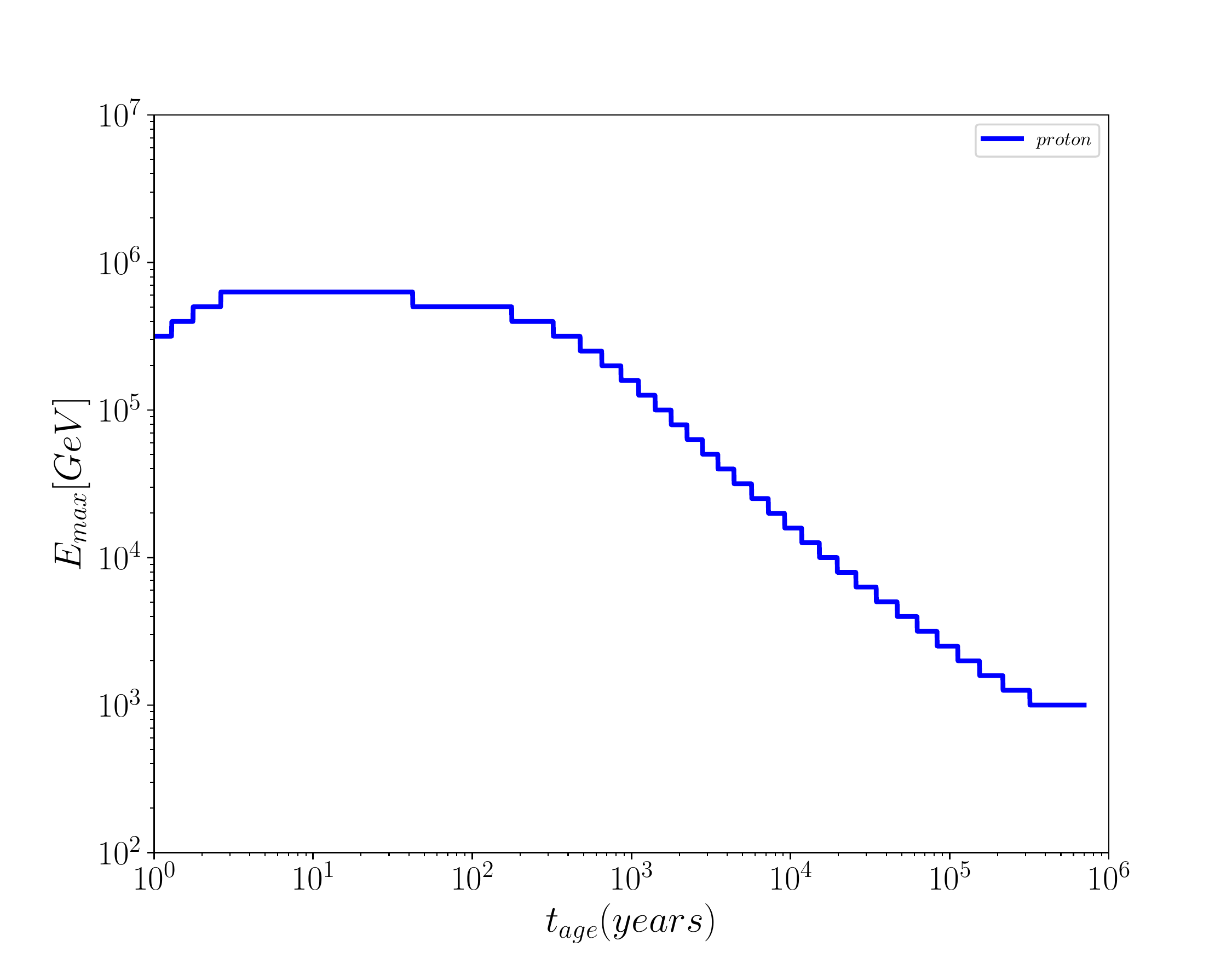}
\caption{The maximum energy of cosmic-ray protons accelerated by the shock inside the nearby SNR.}
\label{figure3}
\end{figure}

\begin{figure*}
    \includegraphics[width=\columnwidth]{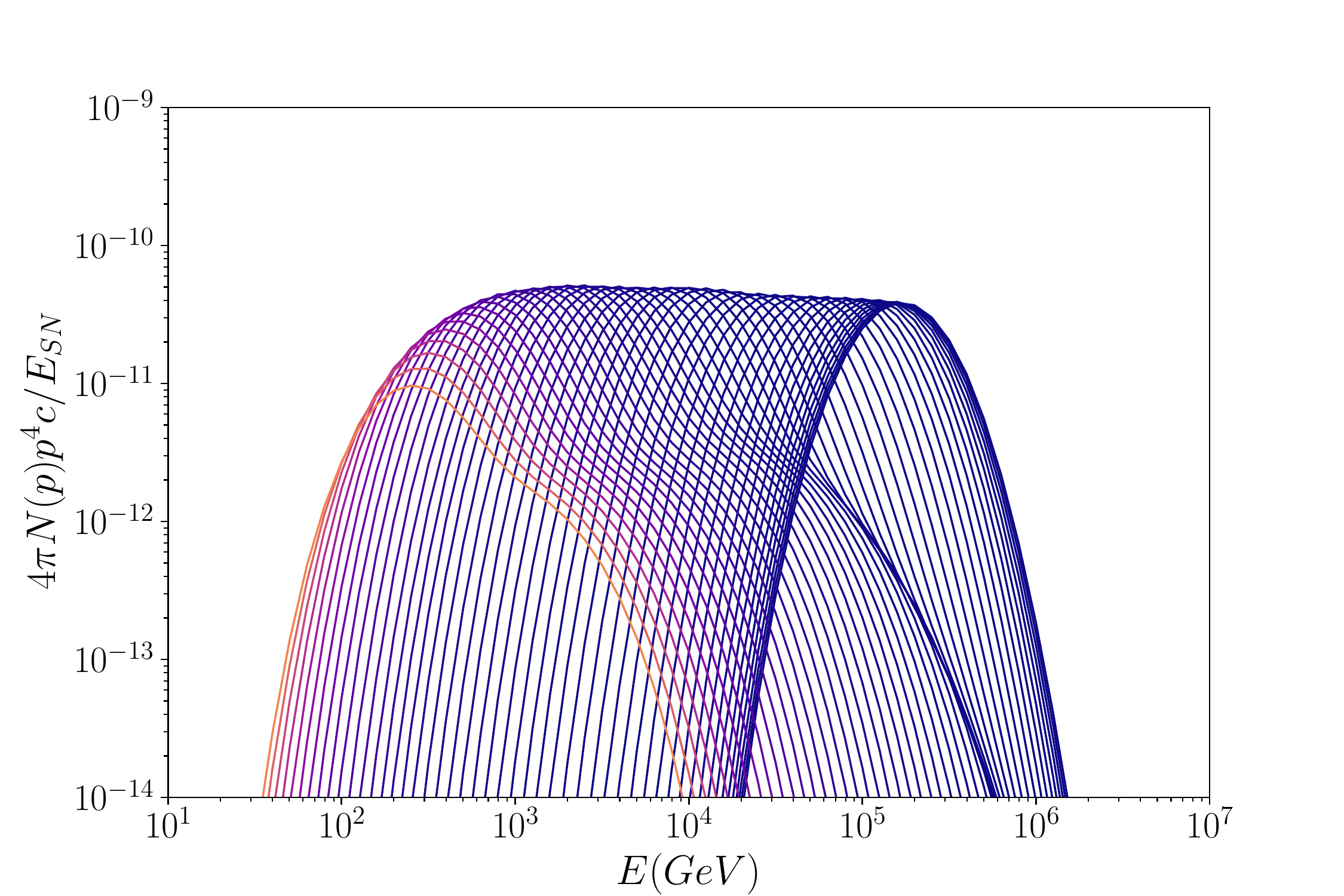}
    \includegraphics[width=\columnwidth]{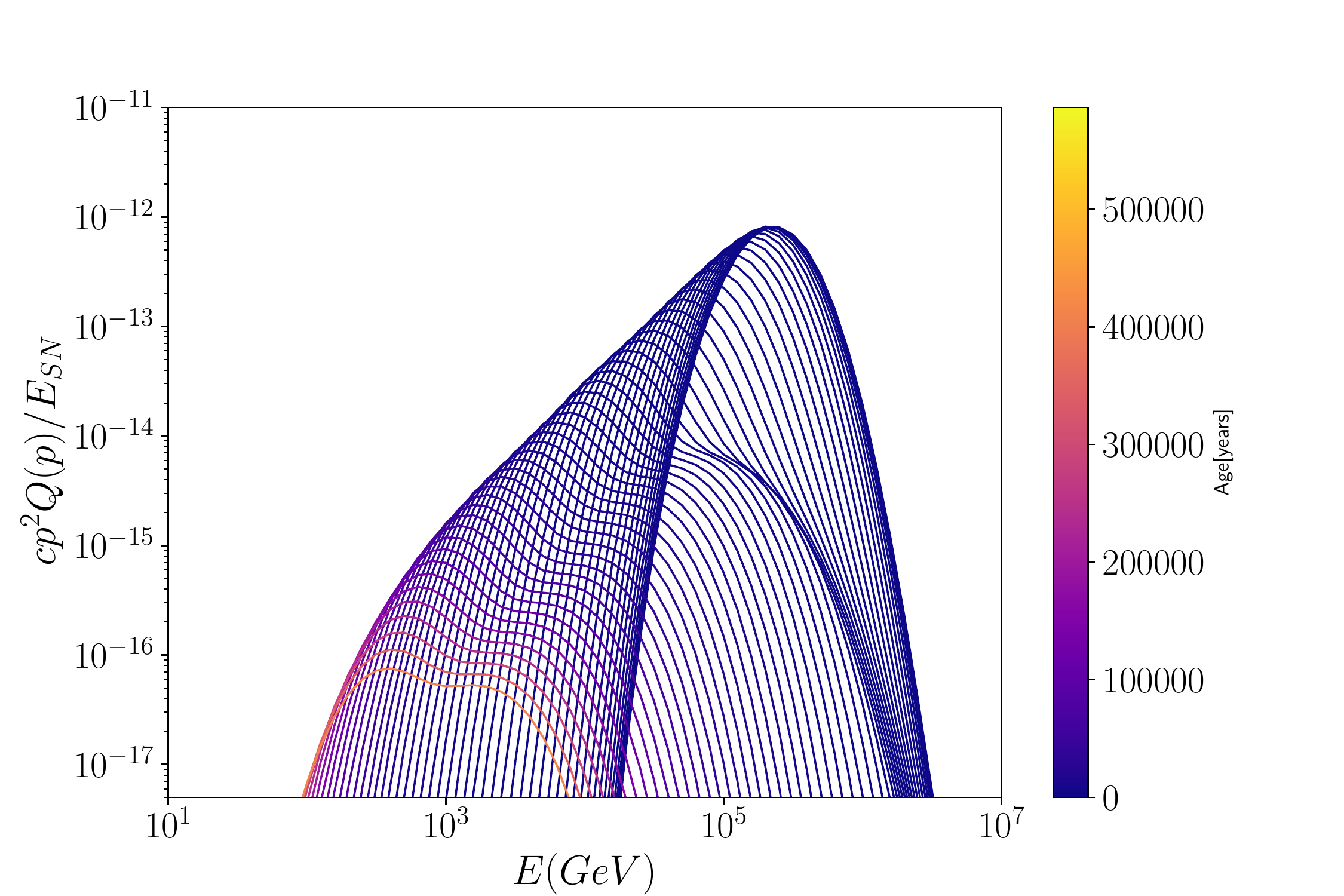}
    \caption{The left panel is the energy distributions of the protons accelerated by the nearby supernova remnant at the around positron of $r=2R_{f}$ of the nearby supernova remnant, which evolves with the SNR age, and the right panel is the distributions of accelerated protons injected in the interstellar medium per second .}
    \label{figure4}
\end{figure*}
We use the numerical package DRAGON to solve the above propagation equation of CRs \citep{2017JCAP...02..015E}. For the Galactic magnetic field structure, it has been selected in the DRAGON code of the type Pshirkov \citep{2011ApJ...738..192P}, in which the disk component is fixed as $\rm B_{0,disk}=2 ~\mathrm{\mu G}$, the halo component has a value $\rm B_{0,Halo}=4 ~\mathrm{\mu G}$, and that of the turbulent component is $\rm B_{0,turb}=7.5 ~\mathrm{\mu G}$.
According to the previous results\citep{2017JCAP...09..029J,2020JCAP...02..009F}, the injection spectra of heavy nuclei and proton are assumed to follow a broken power-law
\begin{equation}
n\left(\rho_{A}\right)=n_{A} \times \begin{cases}\left(\frac{\rho_{A}}{\rho_{b}}\right)^{-\nu_1} ; & \rho_{A} \leq \rho_{b} \\ \left(\frac{\rho_{A}}{\rho_{b}}\right)^{-\nu_2} ; & \rho_{A}> \rho_{b}\\
\end{cases}
\end{equation}
Where $n_{A}$ is injection particle number at the rigidity $\rm \rho_{b}$, and $\nu$ is spectral index.

\section{NEARBY SUPERNOVA remnant} \label{sec:local_source}
The diffusive shock acceleration (DSA) mechanism, believed to occur inside the supernova remnants, predicts a particle spectrum that is in rough agreement with the observed cosmic-ray spectrum corrected for the propagation effects \citep{2020ApJ...894...51A}. After supernova explosion, the supernova ejecta moves super-sonically inside circumstellar medium. This process will generate a forward shock propagating in the circumstellar medium and a reverse shock propagating in the gas of ejecta \citep{2012APh....39...12Z}. The cosmic-ray particles are accelerated by both the forward and reverse shocks. Finally, those cosmic-ray particles escape from the supernova remnants and diffuse in the interstellar medium.

To simulate the acceleration of cosmic-ray particles inside SNRs, the hydrodynamic equations need to be solved together with the diffusion convection transport equation depending on the time $\rm t$, radial distance from the point of supernova explosion $\rm r$ and the particle momentum $\rm p$. The shocks could be modified by the pressure of accelerated cosmic-ray particles and self-consistently modify and determine the spectrum of cosmic-ray particles \citep{2012APh....39...12Z,2010ApJ...708..965Z,2010ApJ...718...31P}. The full process is described by the following equations
\begin{equation}
\frac{\partial \rho}{\partial t}=-\frac{1}{r^{2}} \frac{\partial}{\partial r} r^{2} u \rho
\end{equation}
\begin{equation}
\frac{\partial u}{\partial t}=-u \frac{\partial u}{\partial r}-\frac{1}{\rho}\left(\frac{\partial P_{g}}{\partial r}+\frac{\partial P_{c}}{\partial r}\right)
\end{equation}
\begin{equation}
\frac{\partial P_{g}}{\partial t}=-u \frac{\partial P_{g}}{\partial r}-\frac{\gamma_{g} P_{g}}{r^{2}} \frac{\partial r^{2} u}{\partial r}-\left(\gamma_{g}-1\right)(w-u) \frac{\partial P_{c}}{\partial r},
\end{equation}
\begin{equation}
\begin{aligned}
&\frac{\partial N}{\partial t}=\frac{1}{r^{2}} \frac{\partial}{\partial r} r^{2} D(p, r, t) \frac{\partial N}{\partial r}-w \frac{\partial N}{\partial r}+\frac{\partial N}{\partial p} \frac{p}{3 r^{2}} \frac{\partial r^{2} w}{\partial r} \\
&+\frac{\eta^{f} \delta\left(p-p_{f}\right)}{4 \pi p_{f}^{2} m} \rho\left(R_{f}+0, t\right)\left(\dot{R}_{f}-u\left(R_{f}+0, t\right)\right) \delta\left(r-R_{f}(t)\right) \\
&+\frac{\eta^{b} \delta\left(p-p_{b}\right)}{4 \pi p_{b}^{2} m} \rho\left(R_{b}-0, t\right)\left(u\left(R_{b}-0, t\right)-\dot{R}_{b}\right) \delta\left(r-R_{b}(t)\right)
\end{aligned}
\end{equation}

Here $\rho$ is the gas density, $u$ is the gas velocity, $P_{g}$ is the gas pressure, $m$ is the mass of thermal protons injected at the fronts of forward and reverse shocks at $r=R_{f}(t)$ and $r=R_{b}(t)$ respectively, $P_{c}=4 \pi \int p^{2} d p v p N / 3$ is the cosmic ray pressure, $w(r,t)$ is the advective velocity of CRs, $\gamma_g$ is the adiabatic index of the gas, and $D(r,t,p)$ is the CR diffusion coefficient inside SNRs. $\eta_{f}$ and $\eta_{b}$ represent the injection efficiency at the fronts of forward shock and reverse shock. $R_{f}$ and $R_{b}$ are the forward and reverse shock radius, respectively. It reminds \citep{2010ApJ...708..965Z,2012APh....39...12Z} to have a detailed description of this model.

We assume that the cosmic-ray particles accelerated by the shocks of the SNR are injected into the interstellar medium through an absorbing boundary at position of $r=2R_{f}$ throughout this paper. The diffusion flux of accelerated protons through the boundary of the calculation domain can be described as
\begin{equation}
\frac{\partial F}{\partial t}=4 \pi r^2(-D \partial f / \partial r)
\end{equation}
Therefore, the source function is expressed as $Q(p)=4\pi p^{2}F(p)$.
After escape from the source, the time that released particles travel to the Earth depends on the particle energy. The high-energy particles are accelerated and injected into ISM at the early stages of the SNR evolution \citep{2021PhRvD.104j3013F,2020JCAP...02..009F}. Therefore, we consider the injection as a decaying injection. It can be described through the following diffusion-loss equation that cosmic-ray protons from a nearby supernova remnant inject into the interstellar medium and diffuse from the nearby source to the Earth
\begin{equation}
\frac{\partial f(E, t, r)}{\partial t}=\frac{D(E)}{r^{2}} \frac{\partial}{\partial r} r^{2} \frac{\partial f}{\partial r}+\frac{\partial}{\partial E}(b(E) f)+Q(E,t)
\label{diffusion-loss}
\end{equation}
Where $Q(E,t)$ is the injection term of cosmic-ray protons from the nearby SNR. $b(E)$ is the rate of energy loss.
Here, we adopt a model where the spatial diffusion coefficient is correlated with the matter distribution. The spatial diffusion coefficient has an energy dependence, as the damping of diffusion depends on the local environments \citep{2014ApJ...782...36E}. Therefore, the energy dependence of the diffusion coefficient for CRs is mainly affected by the different magneto-hydrodynamic (MHD) properties along the Galactic interstellar medium. The CRs released by the SNR may amplify the magnetic field turbulence of the ambient medium through the streaming instability \citep{1971ApJ...170..265S}, which may lead to slow diffusion in the vicinity of the nearby SNR. Therefore, we consider the diffusion coefficient as a two-zone diffusion model in which the diffusion is slow in a small region around the source, out of which the propagation is as fast as usual. This scenario is consistent with two diffusive zones proposed in \cite{2012ApJ...752L..13T}, in which the parameters ($\xi$,$\Delta$) are used to characterize a smooth transition of the diffusion coefficient in the two zones (inner and outer halos). The limit of $\xi\to 1$ or $\Delta\to 0$ will make the two-zone diffusion effect pass to the homogeneous or one-zone diffusion.
We adopt the diffusion coefficient with a slope depending on the energy
\begin{equation}
D(E)=D_{0}\left(\frac{E}{E_{0}}\right)^{\gamma(E)}
\end{equation}
where $\rm D_0\approx1.21 \times 10^{28} ~cm^{2}s^{-1}$, and $\gamma(E)$ is take the setup considered in \citep{2012ApJ...752L..13T}. It is assumed as
\begin{equation}
\gamma(\rho) \approx \gamma_{\text {high }}+\frac{\Delta}{1+\frac{\xi}{1-\xi}\left(\frac{\rho}{\rho_{0}}\right)^{\Delta}}
\end{equation}
where $\rho$ is particle rigidity, parameters are taken the values $\gamma_{high}=1/6$, $\Delta=0.55$, $\xi=0.1$ and $\rho_0=2~\mathrm{GV}$.
With the source injection spectrum simulated in the nearby SNR and maximum energy of protons accelerated inside that at evolution time $\mathrm{t}$, we solve the above diffusion-loss Eq.\ref{diffusion-loss} to calculate the proton density per unit energy reaching the Earth.

\section{RESULTS AND DISCUSSION}
\label{sec:result}
\begin{figure}[t]
\centering
\includegraphics[width=\linewidth]{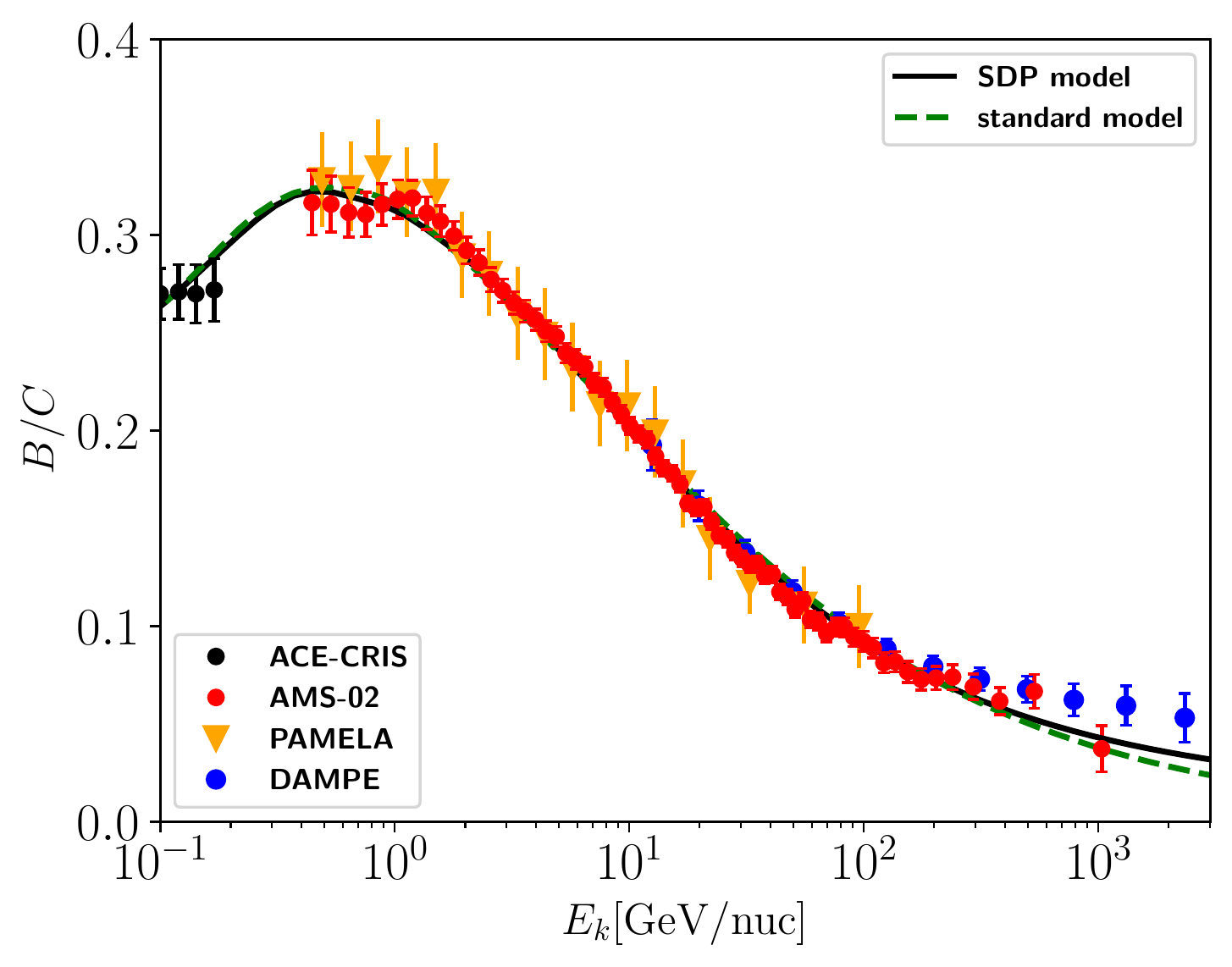}
\caption{The B/C ratio computed for the standard cosmic-ray propagation model and spatial dependent model, compared with the experimental AMS-02 \citep{2016PhRvL.117w1102A}, PAMELA \citep{2014ApJ...791...93A}, ACE-CRIS \citep{2019SCPMA..6249511Y} and \citep{2022SciBu..67.2162D} data.}
\label{figure5}
\end{figure}
\begin{figure}[t]
\centering
\includegraphics[width=\linewidth]{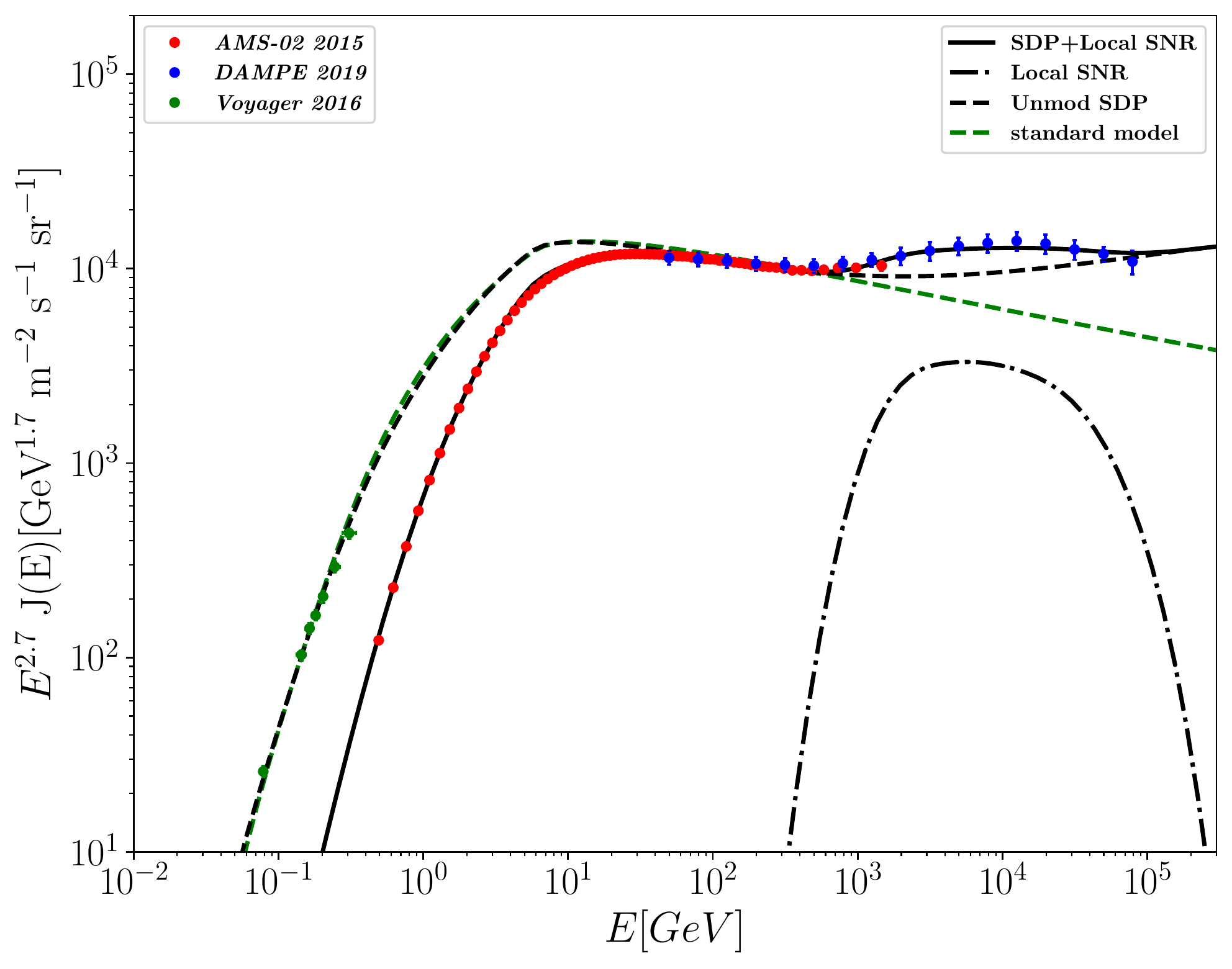}
\caption{The black solid line represents the total proton spectrum which is the sum of the large-scale proton spectrum calculated by the spatial dependent cosmic-ray propagation model and the component provided by the nearby supernova remnant , and the black dashed line shows the un-modulated background proton spectrum calculated by the SDP. The green dashed line represents background CR proton spectrum calculated by traditional model, and the black dashed-dotted line is the proton spectrum from the nearby supernova remnant. The proton flux data from AMS-02 \citep{2015PhRvL.114q1103A}, Voyager \citep{2016ApJ...831...18C}, and DAMPE \citep{2019SciA....5.3793A} are also shown.}
\label{figure6}
\end{figure}
\begin{figure}[t]
\centering
%\begin{minipage}{\textwidth}
\includegraphics[width=\linewidth]{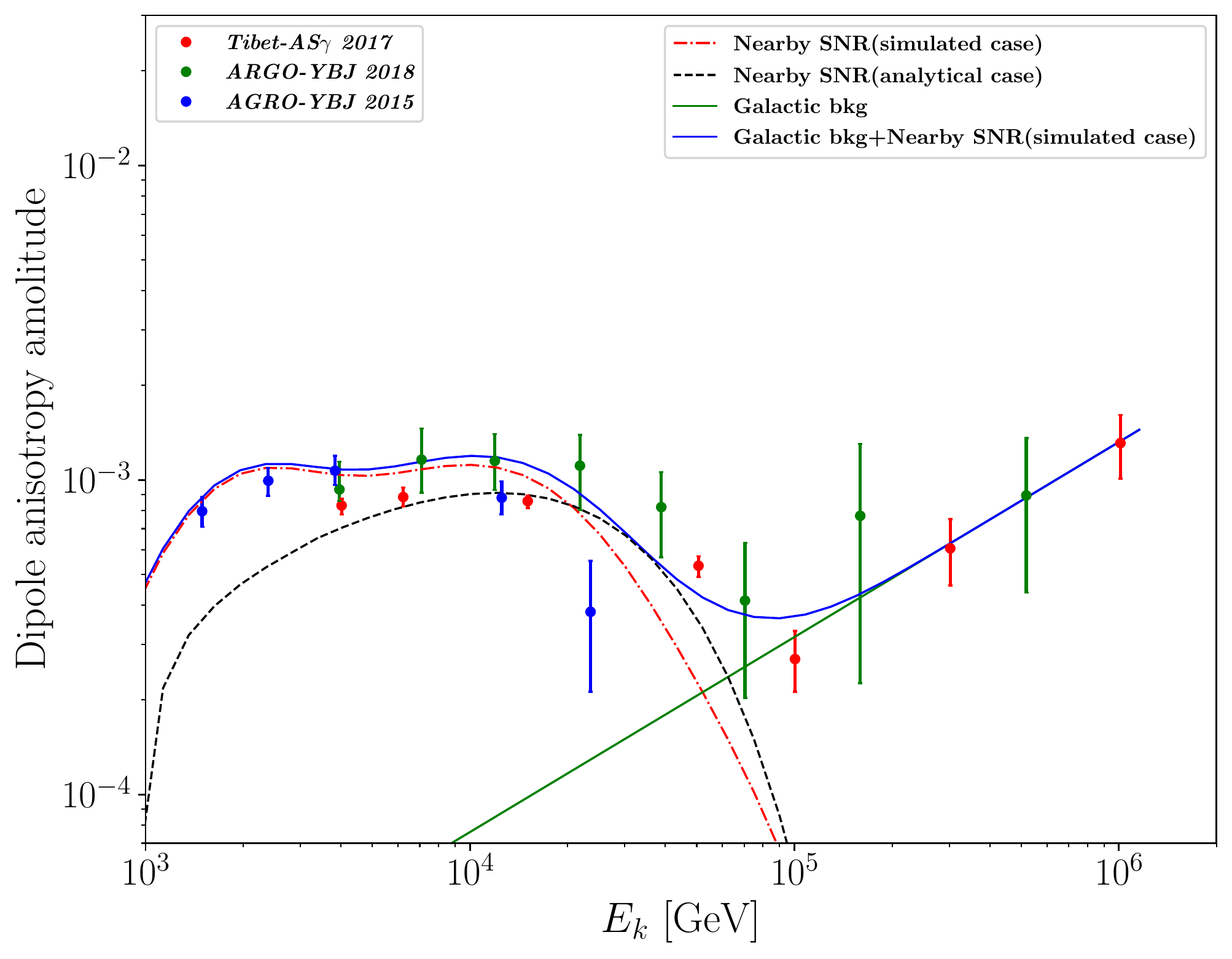}
%\end{minipage}
\caption{The cosmic-ray dipole anisotropy with the background (green solid line) and nearby supernova remnant (red dashed-dotted and black dashed lines) contribution calculated by the model, compared with the data from Tibet-AS$\gamma$ \citep{2017ApJ...836..153A} and ARGO \citep{2015ApJ...809...90B,2018ApJ...861...93B}. Here, the red dashed-dotted line is the dipole anisotropy amplitude calculated in our simulation scenario, and the the black dashed line shows one obtained through the analytical case.}
\label{figure7}
\end{figure}
We have modeled the cosmic-ray proton flux data from the AMS-02 and DAMPE experiments using a Galactic population of conventional cosmic-ray nuclei sources and a nearby SNR providing the extra proton component. The DRAGON numerical code was used to calculate the propagation of Galactic cosmic rays.
Besides, to implement a more realistic injection spectrum from the nearby SNR, we have also simulated the acceleration process of cosmic-ray protons inside the nearby SNR, and then calculated the proton fluxes reaching the Earth from the nearby SNR through the diffusive loss equation.

We assume those cosmic-ray particles with a radius $r\geq 2R_{f}$ have  escaped from the supernova remnant and started to inject into the interstellar medium. Besides, we consider the nearby supernova remnant as the Type $\rm \uppercase\expandafter{\romannumeral1}$a supernova which evolves in a constant density medium with a hydrogen number density $\rm n_0=0.01~cm^{-3}$, magnetic field strength $\rm B_0=5~\mu G$ and temperature $\rm T=10^4~K$. After the supernova explosion, the eject mass and explosion energy are considered as $1.5~M_{\odot}$ and $\rm 2.7\times 10^{51}~erg$, respectively. The injection efficiency $\eta_f=\eta_b=0.01$ was used to modify shocks inside SNR.
We show the evolution of some physical quantities depending on supernova remnant radius in Fig.\ref{figure1}, such as gas density, gas velocity, cosmic-ray pressure, and gas pressure. As the simulated results, the radius and velocity of the supernova remnant forward and reverse shocks, and the magnetic field strength downstream of forward shock are also shown in Fig.\ref{figure2}.
The results simulated from the nearby supernova remnant in this work almost agree with the previous productions (see the \citep{2012APh....39...12Z,2010ApJ...708..965Z}). To calculate the proton flux reaching the Earth from the nearby supernova remnant, we also simulated the maximum energy of cosmic-ray protons accelerated inside SNR (see Fig.\ref{figure3}) and the proton spectrum evolving with the time (see Fig.\ref{figure4}) at positron $r=2R_{f}$. We can find that the spectrum of runaway protons moves from the high energy band to the low energy band with the evolution time. The reason may follow that the magnetic field and shock velocity both decrease gradually with evolution time. While the magnetic field influences high-energy protons through the efficiency of confinement around the forward shock \citep{2004MNRAS.353..550B,2010ApJ...718...31P,2022ApJ...926..140S}. Therefore, the proton maximum energy accelerated in SNR is decreasing gradually with time due to the adiabatic energy loss and the decrease of the magnetic field strength.

\begin{table*}
\begin{center}
\caption{The parameters of propagation and injection spectrum.}
\footnotesize
\setlength{\tabcolsep}{3mm}{
\begin{tabular}{cccccccccccc}
\label{tab:1}\\
\hline\hline
 Model & $ \nu_{1} $ & $\nu_{2}$& $\rho_{b}(\rm GV)$ & $ v_{A}(kms^{-1}) $ & $\delta$ & $D_{0}(cm^{2}s^{-1})$ & $\rho_{0}(\rm GV)$ & $\eta$ & $N_{m}$ & $n$ & $\xi$\\
\hline
standard & $1.9$ & $2.4$&$7$& $18$ &$0.45$ & $2.25$ & $1.0$ & $0.2$ & $-$ & $-$ & $-$\\
\hline
SPD & $1.9$ & $2.4$& $7$&$5$ &$0.6$ & $4.8$& $4.5$ & $-0.4$ & $0.24$ & $5$ & $0.092$\\
\hline
\end{tabular}}
\end{center}
\end{table*}

For the background cosmic-ray propagation, we use the Boron-to-Carbon ratio to determine the parameters in diffusion coefficient. We fix the halo size $\rm H=4~kpc$ \citep{2023FrPhy..1844301M} and then obtain $\delta$. The Alfven velocity is also fixed through tuning to the low-energy data (e.g., ACE-CRIS data \citep{2019SCPMA..6249511Y}). The corresponding model parameters are shown in Table.\ref{tab:1}. In Fig.\ref{figure5}, we show the B/C ratio results calculated by the cosmic-ray propagation model and observed data provided by the AMS \citep{2016PhRvL.117w1102A} and PAMELA experiments \citep{2014ApJ...791...93A}. The results show that the spatial-dependent re-acceleration CR propagation scenario produces a hardening at $\sim \rm 200~GV$, compared with the traditional CR propagation model. This may be due to the effect of the change in the turbulence properties of ISM. But it may not seem to reproduce hardening of B/C reported recently by the DAMPE Collaboration \citep{2022SciBu..67.2162D}. It is believed that the interactions of primary CRs with the ISM may occur in the vicinities of CR sources, and secondary particles generated close to the accelerating sources may have chance to be accelerated by the shocks of the sources \citep{2009PhRvD..80f3003F,2016PhRvD..94f3006M,2019PhRvD.100f3020Y}, resulting in a harder spectra of secondary particles. Such interactions can provide contributions for the secondary CR particles and explain the current DAMPE data \citep{2009PhRvD..80l3017A,2009PhRvL.103h1104M,2022arXiv221009205M}. The detailed interpretation of the features of B/C fluxes detected recently by the DAMPE is far beyond the scope of this paper, and we will address it in a future work.

In Fig.\ref{figure6}, we show the proton spectrum including the diffusion background cosmic-ray component and the contribution from the local supernova remnant. The lower energy band cosmic-ray particles would be influenced by the solar modulation. Here, Voyager data (below $\rm 1 ~ GeV/n$) \citep{2016ApJ...831...18C} outside the Heliosphere are used to tune the low-energy CR spectra that have not been affected by solar modulation. Besides, we consider this modulation as the force field approximation \citep{1968ApJ...154.1011G} which has an effective potential $<\phi_{mod}>$. We constrain the AMS-02 data \citep{2015PhRvL.114q1103A} to get the $<\phi_{mod}>\sim 250 \rm ~ MV$.
As shown in Fig.\ref{figure6}, the observed spectra are reproduced by introducing contributions of the background cosmic rays calculated by the spatial-dependent CR propagation model and that from a nearby source. According to the constrain of observed data, the nearby supernova remnant may lie at a distance of $\rm \sim 350~pc$ and have an age of $\rm \sim 1\times10^{5}~yr$. We find that the nearby-source contribution is crucial for the interpretation of the ``bump" structure of the cosmic-ray proton spectrum reported by the DAMPE Collaboration. The position of the peak in the proton flux spectrum ($\sim \rm 10~TeV$) requires a released time $\rm t_{rel}\sim10000~yr$ to match the observations. Because SNR will release particles with higher energies at an early stage (see Fig.\ref{figure4}), while the low-energy component is released at a time close to the current age of SNR. Therefore, at the earlier epochs of SNR evolution, the presence of high-confinement regions around this nearby source has to be required to match the observed data. In fact, when the SNR encounters the dense Molecular Clouds clump, the shock is rapidly stalled \citep{2010ApJ...724...59S,2016EPJWC.12104001G}. The shell of amplified magnetic turbulence is formed at the stalled shock after the shock-cloud collision, and the release of the CRs may not be an immediate event \citep{2012ApJ...744...71I}. This turbulent shell could last for $\sim \rm 10^3 yr$ when a fast shock is hitting the Molecular Clouds clump. The relative earlier epochs of SNR would need to take this amplified magnetic field shell into consideration, and this effect is simplified as a delayed releasing time of the CRs after the shock-cloud collision.

However, the other primary CR particle spectrum has also a observed break at $\sim 300 \rm ~GV$ \citep{2017PhRvL.119y1101A,2018PhRvL.120b1101A}. Although the local source scenario could also provide some contributions, this effect could originate mainly from the CR propagation mechanism (e.g., spatially dependent diffusion), which can lead to a larger spectral break of the secondary species with respect to primaries. This steeper spectrum of the secondaries have been revealed \citep{2021PhR...894....1A}. The Galactic cosmic-ray flows can induce MHD waves of the background plasma \citep{1971ApJ...170..265S}, leading to self-confinement of Galactic cosmic rays around such waves. This nonlinear effect results in changes in the momentum dependence and spatial dependence of the diffusion coefficient, which is a possible mechanism of the hardening feature \citep{2012PhRvL.109f1101B}. Therefore, the harder high energy part of the B/C ratio or the hardening at $\rm \sim 300~GV$ of protons may be due to a transition of the diffusion from self-generated turbulence to externally generated turbulence \citep{2023FrPhy..1844301M}. This is the reason why we use a spatial-dependent propagation model to calculate the background cosmic-ray component.
\begin{figure*}
\includegraphics[width=\columnwidth]{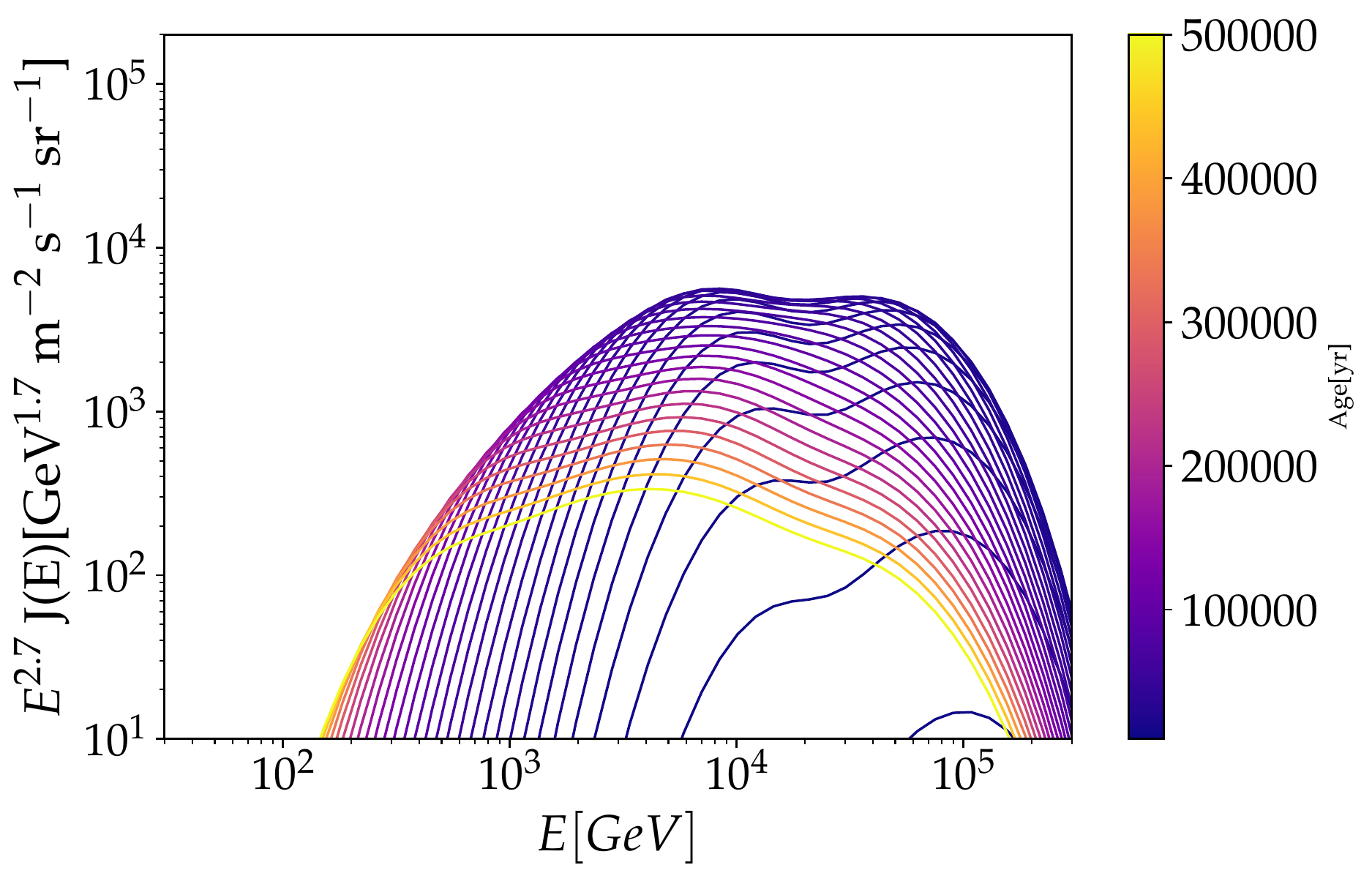}
\includegraphics[width=\columnwidth]{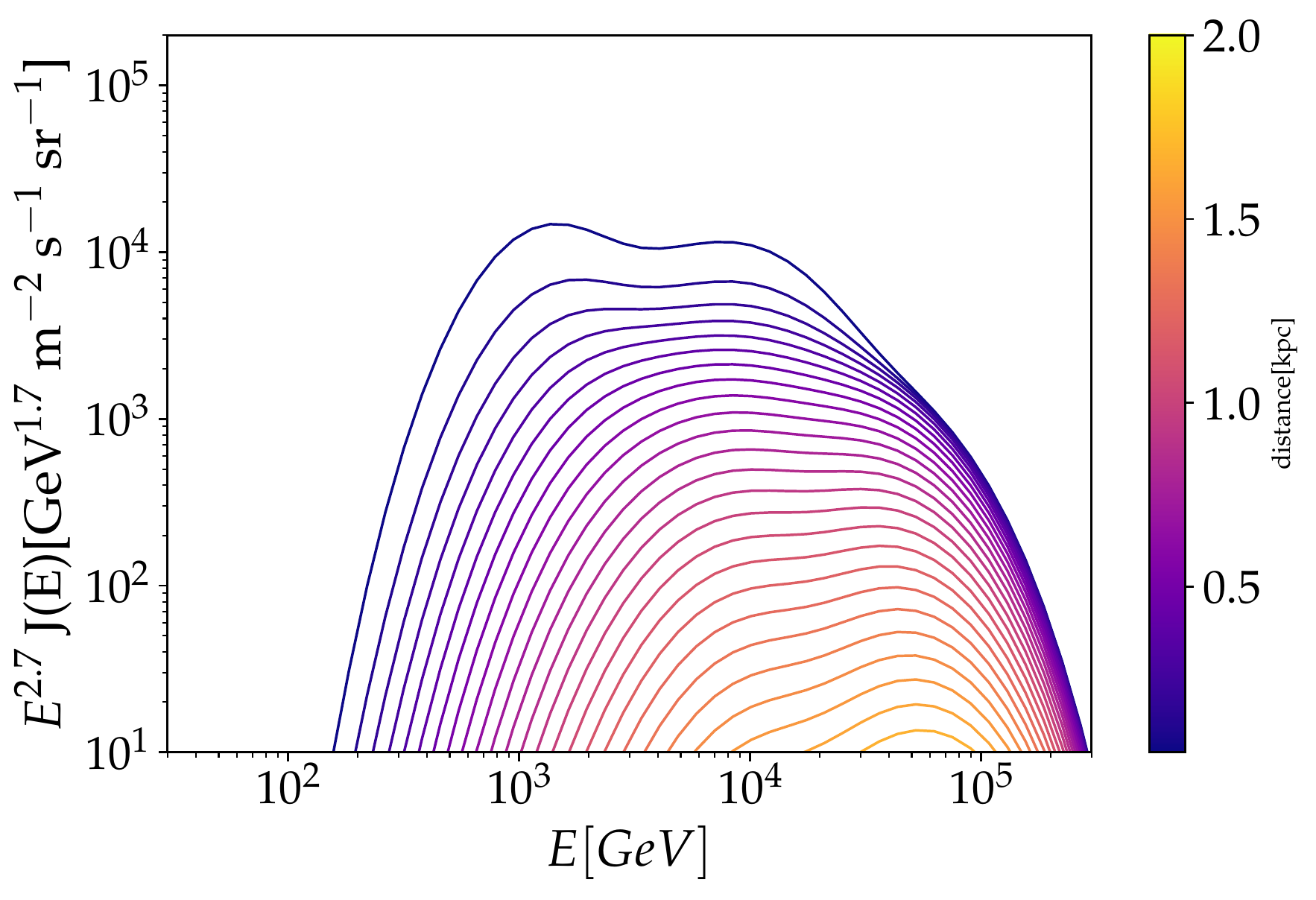}
\includegraphics[width=\columnwidth]{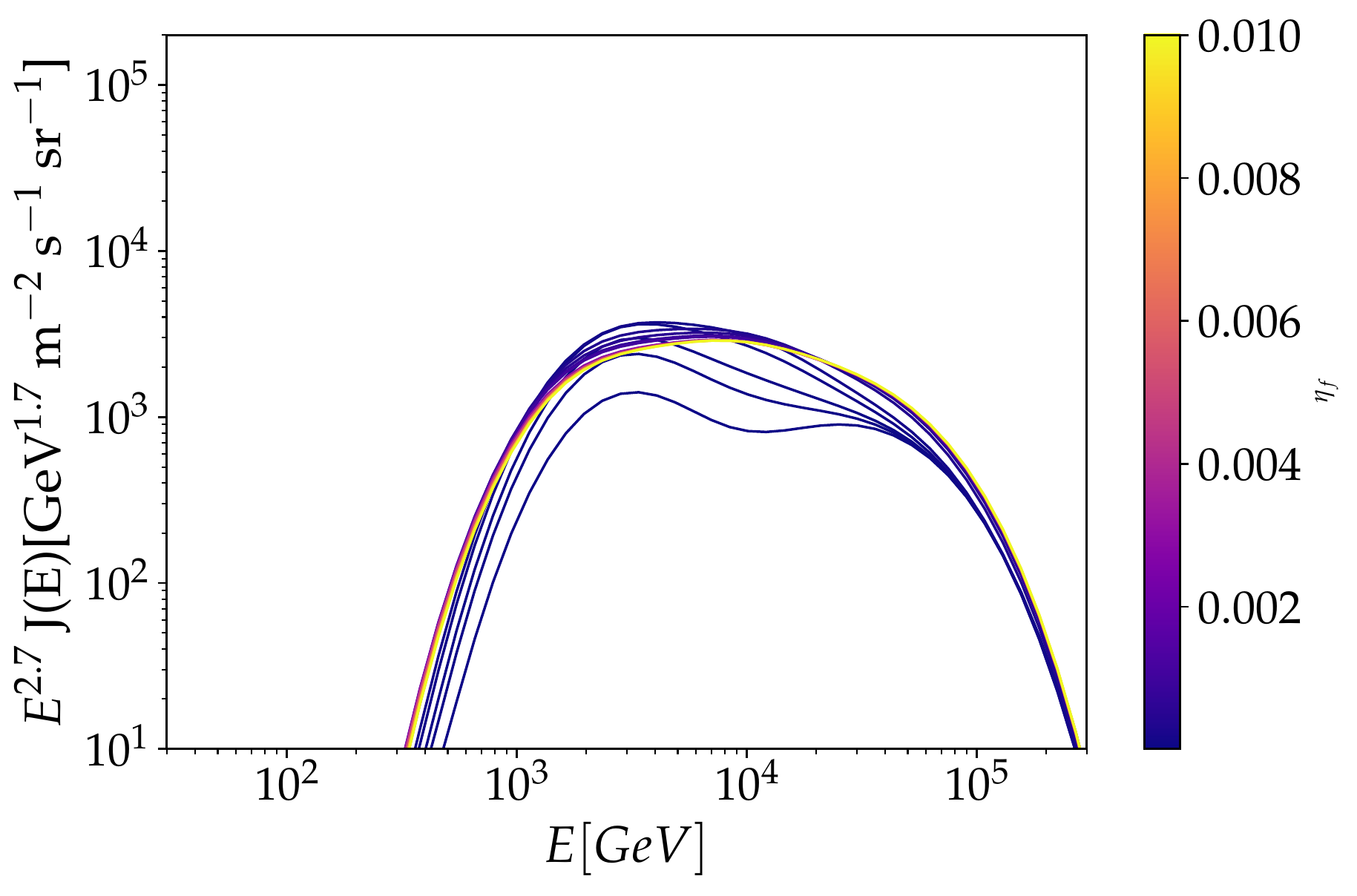}
\includegraphics[width=\columnwidth]{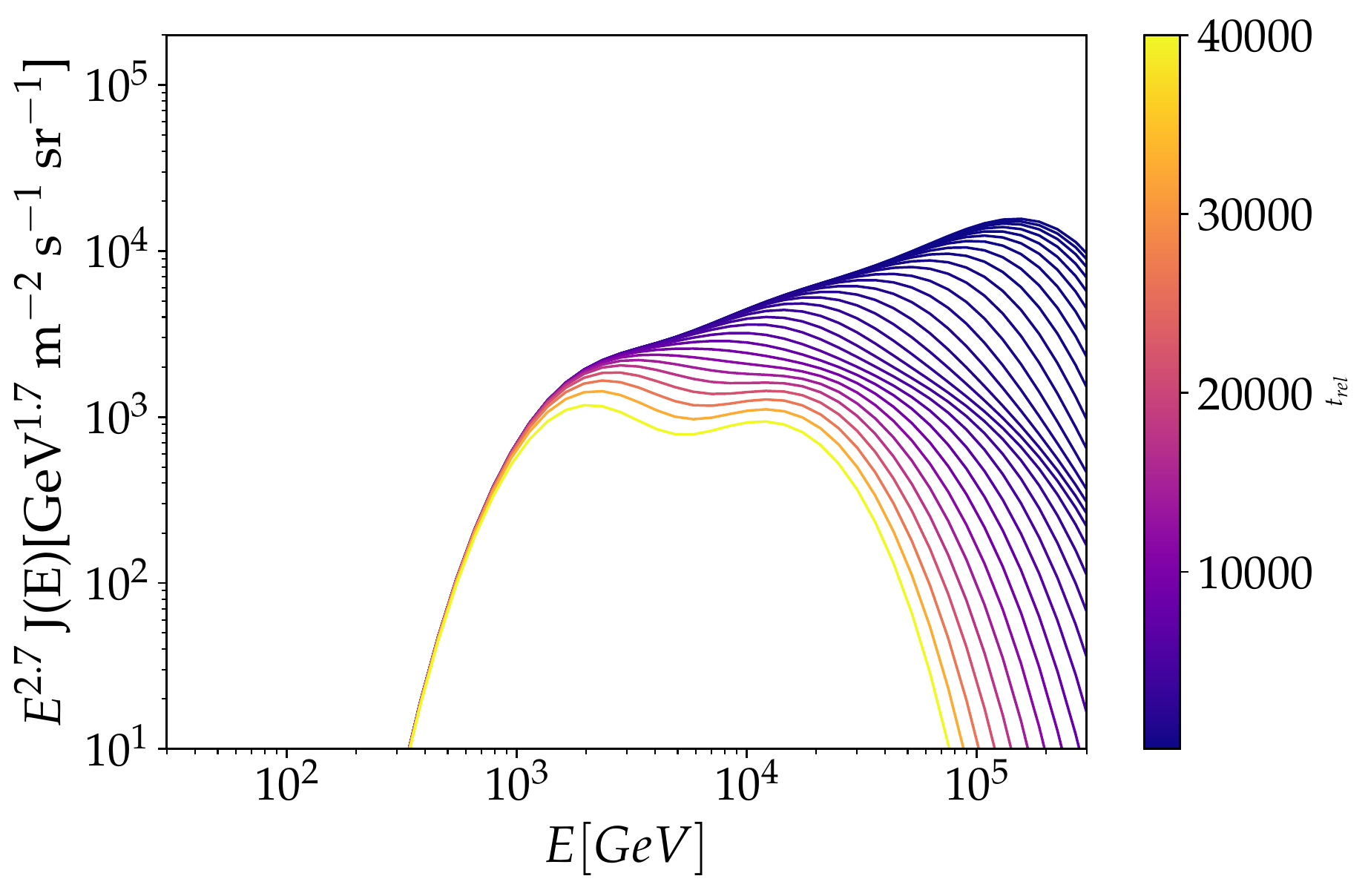}
\caption{Examples of the spectral influence of injection efficiency, release time of CR protons, and the age and distance of local SNR. The panel in the figure shows the spectra of CR proton fluxes of the nearby SNR. The different curves represent the spectrum calculated by values of parameters given by the color scale, keeping the other parameters in each panel fixed at the fitting value adopted by Fig.\ref{figure6}. }
\label{figure8}
\end{figure*}

The cosmic-ray dipole anisotropy is crucial for constraining the model. The Fig.\ref{figure7} shows amplitudes of the dipole anisotropy predicted by the model, compared
with the observed data. We also show the background CR anisotropy using results fitting in \cite{2020ApJ...903...69F} (see the green line of Fig.\ref{figure7}). Here, the dipole anisotropy is calculated by the $\frac{3 D_{x x}}{v} \frac{|\nabla \psi|}{\psi}$, the background anisotropy is written as $\Delta_{\mathrm{bkg}}=c_{1}\left(\frac{E}{1 \mathrm{PeV}}\right)^{c_{2}}$ and $c_1=1.32\times10^{-3}$, $c_2=0.62$. It is interesting to note that the anisotropy amplitudes below the $\rm \sim 100 ~ TeV$ energy band show a difference of about two orders of magnitudes between the background component and observed data. In previous works, it is explained that the nearby source has a relatively high CR contribution in this energy range and dominates the total anisotropies of CR particles \citep{2019JCAP...12..007Q,2020FrPhy..1624501Y}. According to the proton fluxes accelerated in the nearby supernova remnant, we calculate the contribution of anisotropy from nearby SNR in our work. As shown in Fig.\ref{figure7}, in the around $\rm 10-100~TeV$ energy band, the spectra of the cosmic-ray dipole anisotropy from the nearby supernova remnant is compatible with the current anisotropy data. The anisotropy from the local source dominates the CR anisotropy features in this energy band. (see the red dashed-dotted line in Fig.\ref{figure7}). In Fig.\ref{figure7}, We also added the dipole anisotropy spectrum estimated in an analytical way to compare both results. For the same basic source parameters, both results have significantly different. The spectrum obtained with our simulation case is consistent with the observed data well. But the observed data is flatter, compared with the spectrum calculated by the analytical formula. Besides, the phase of dipole anisotropy at $E\lesssim 100 \rm ~TeV$ points to the direction of the anti-Galactic center, and turns to the direction of Galactic center at $E\gtrsim 100 \rm ~TeV$ according to the observed results \citep{2017PrPNP..94..184A}. This is because that local source contribution dominants the dipole anisotropies at $E\lesssim 100 \rm ~TeV$, which keeps the direction of the local source, and that at $E\gtrsim100 \rm ~TeV$ is dominated by the background component instead, which points to the Galactic center direction due to the more abundant CR sources.
\begin{figure*}
    \includegraphics[width=\columnwidth]{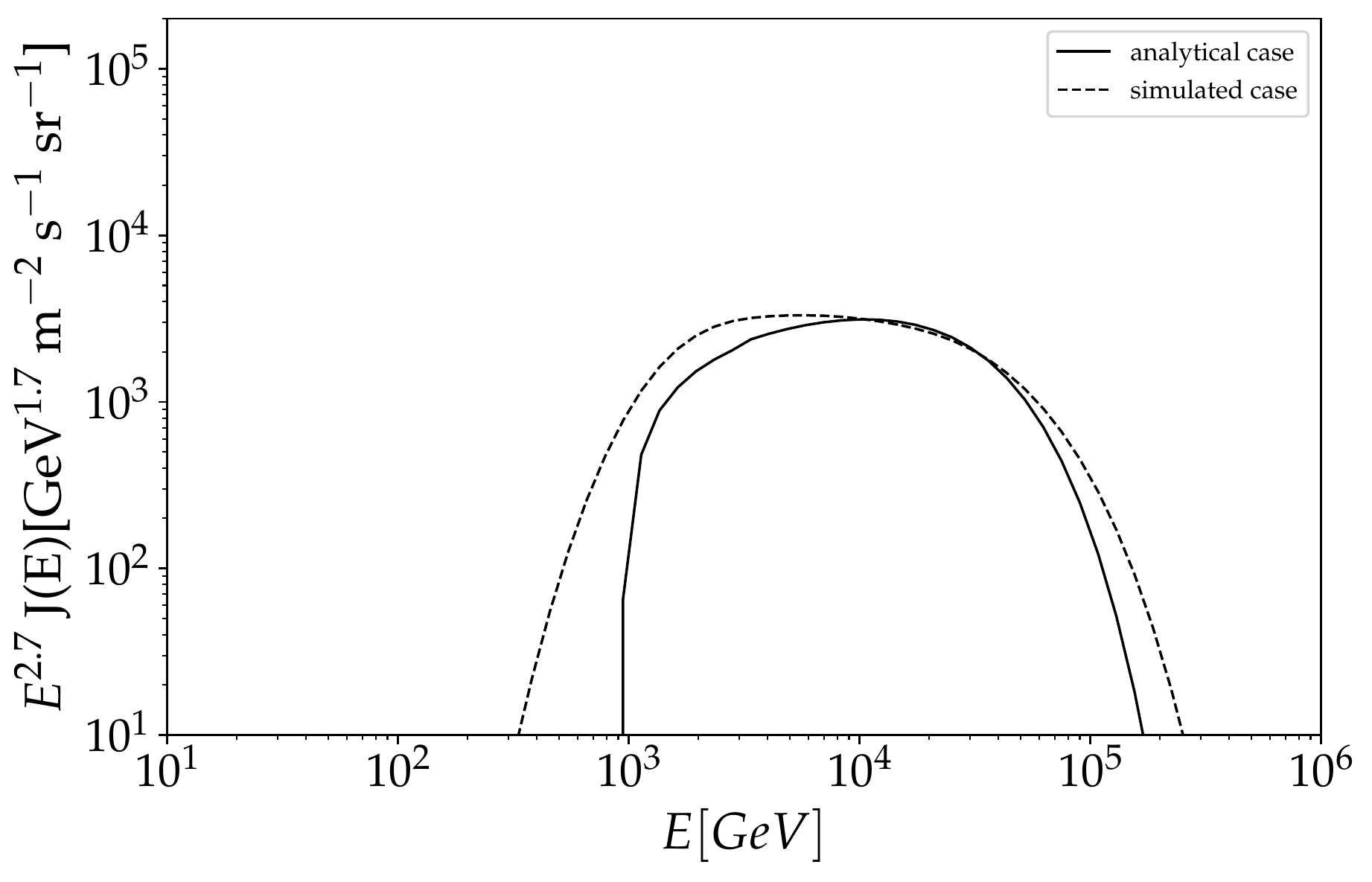}
    \includegraphics[width=\columnwidth]{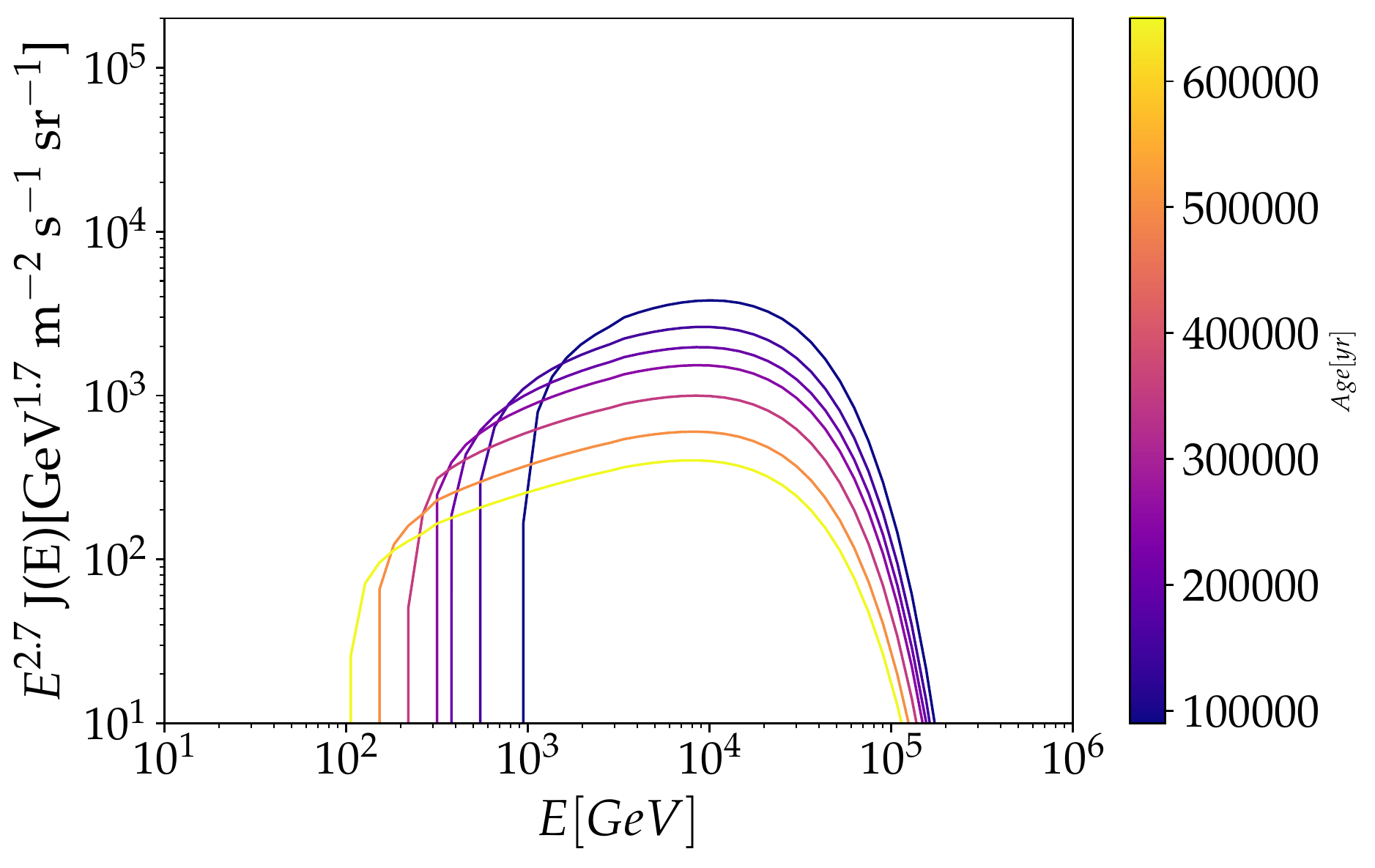}
    \includegraphics[width=\columnwidth]{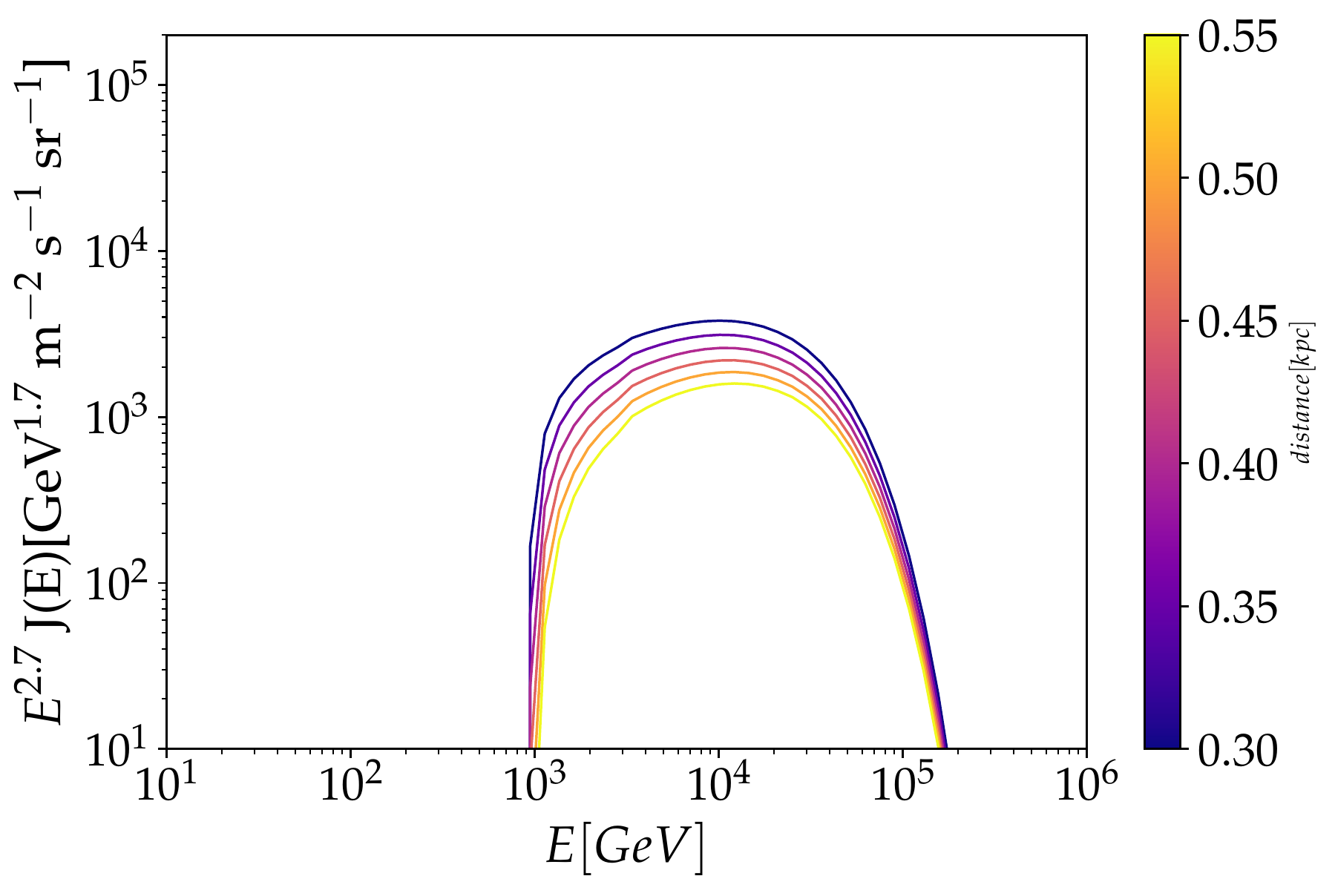}
    \includegraphics[width=\columnwidth]{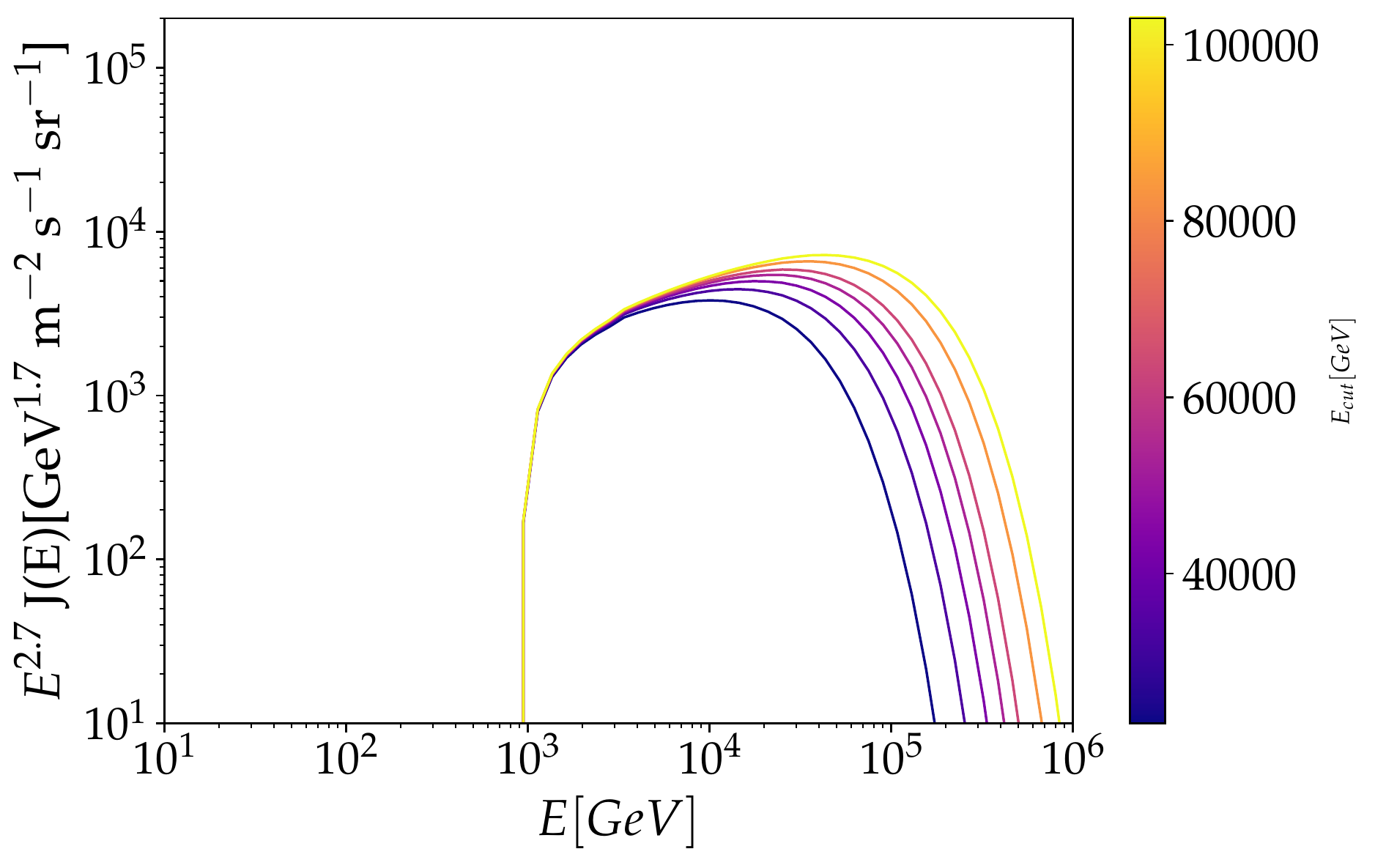}
    \caption{The top left panel shows the comparison between the flux of protons from the local SNR calculated with the simulation way and the one obtained from the analytical scenario. The other panels show examples of the spectral influence of break energy $\rm E_{cut}$, and the age and distance of local SNR in an analytical scenario. }
    \label{figure9}
\end{figure*}

For the analytical injection scenario, the resulting spectral shape is almost invariable. The age of the nearby SNR mainly affects the spectrum in low energy bands, and the corresponding distance is tuned to match the observed data \citep{2021PhRvD.104j3013F,2019JCAP...10..010L}. Here the runway energy of particles inside SNR is approximately considered as linear decrease with time. The peaking structure of the calculated spectrum is decided by the free parameters $\rm E_{cut}$, while the effect of the energy-dependent release cuts off the low-energy part of the spectrum. It can be found that the peak structure here is caused by mathematical reasons. But in fact, what we are more concerned about is the physical origin of this structure. Therefore, when we use the theoretical spectrum calculated in this way to fit the observed data to further address the origination of this spectral structure or to determine the properties of local SNR, it has strong uncertainty.
In our modeling injection scenario, the corresponding free parameters include the hydrogen number density $\rm n_H$, magnetic field strength $\rm B_0$ and temperature $\rm T$ in the circumstellar medium, the ejecta mass $\rm M_{ej}$, the energy of the explosion $\rm E_{SN}$, and injection efficiency. But the spectral shape of runaway particles mainly depends on the injection efficiency in forward shock. Together with the diffuse process from the nearby source to the Earth, the resulting proton flux spectrum mainly depends on the injection efficiency, age, release time of the particles, and distance of SNR. Examples of how the spectra are affected by the contrast parameters are shown in Fig.\ref{figure8}. It can be found that some physical effects, such as the nonlinear response of energetic particles and the release time of CR protons can leave strong signatures in the spectrum of the resulting CR proton fluxes.

In Fig.\ref{figure9}, we show the comparison of our results with the analytical scenario. We find that both estimations are very similar when we tune the free parameters to match the observed data. However, It is previous that the spectral shape of proton flux obtained in an analytical scenario is almost invariable. This could be because the injection spectrum in the analytical scenario is assumed as the form of $Q(E,r,t)=S_0(\frac{E}{E_0})^{-\Gamma_{inj}}e^{E/E_{cut}}\delta(r)L(t)$, while the corresponding source parameters are only used to estimate proton escape energy as a function of time (see, e.g., \citep{2021PhRvD.104j3013F}). Different from the analytical case, in our simulated scenario, the spectral shape of proton flux from the nearby SNR depends on the source properties, such as the injection efficiency, age, and release time of particles. Therefore, for some situations which have been shown in the other panels of Fig.\ref{figure8}, both scenarios are strongly different (see Fig.\ref{figure8} and \ref{figure9}).

Finally, the recent spectral measurement of helium showed that the drop-off starts from $\rm \sim 34~TV$. Together with the softening energy of the proton spectrum at $\rm \sim 13.6~TV$, both results are consistent with a charge-dependent softening energy of protons and helium nuclei \citep{2021PhRvL.126t1102A}. On the other hand, compared with the proton spectrum, the spectrum of accelerated energetic helium in supernova remnants has the same shape \citep{2010ApJ...718...31P}. Therefore, it seems that the nearby source model also becomes natural and accessible for explaining the DAMPE ``bump" observed for He at almost the same energy per nucleon.

\section{CONCLUSIONS}
\label{sec:conclusion}
The CR proton spectrum reported by the DAMPE Collaboration \citep{2019SciA....5.3793A} and AMS-02 \citep{2015PhRvL.114q1103A} shows a complex ``bump" spectral feature at $ \sim \rm 200~GeV- 10~TeV$ energy band, which is inconsistent with results predicted by the standard cosmic-ray propagation model. The local source is universally considered to explain this phenomenon \citep{2021PhRvD.104j3013F,2020FrPhy..1624501Y,2020ApJ...903...69F}. In this work, we simulated the acceleration of cosmic-ray protons in a local supernova remnant. Then they inject into the interstellar medium at positron of $r=2R_f$ and diffuse to our Earth. The simulated results about the nearby SNR are almost consistent with the previous work \citep{2012APh....39...12Z,2010ApJ...708..965Z}.

According to the simulated results, the sum of component predicted by background cosmic-ray propagation model and contribution of nearby SNR can interpret AMS-02 and DAMPE data. We conclude that the complex proton spectral structure of hardening at $\rm \sim 200~GeV$ and softening at $\rm \sim 10~TeV$ may be dominated by the superposition of the background CR component and nearby supernova remnant contribution. The anisotropy
amplitude is sensitive to the relative flux differences between the background component and the nearby source component, the protons from the nearby SNR dominate the total anisotropies of CR particles in the low energy range. On the other hand, we find that this nearby SNR has a time delay between the supernova explosion and the release of particles in the ISM. This means that those particles accelerated before the release time were not immediately injected into the interstellar medium. Therefore, the presence of high-confinement regions around this nearby source has to be required to match the observed data at the earlier epochs of SNR evolution. While some physical effects, such as the nonlinear response of energetic particles and the release time of CR protons can leave strong signatures in the spectrum.

It should be remarked that the nearby supernova remnant discussed in this work is not a known specific source. Therefore, it needs to make further observations to provide constraints or evidence in future work.

\section*{Acknowledgements}
We thank Zhen Cao and Felix Aharonian for helpful discussion. This work is partially supported by the National key research and development program 202301AS070073, and the National Natural Science Foundation of China (NSFC U1931113, 12233006).

\bibliographystyle{aasjournal}
\bibliography{MS}

\clearpage
\end{document}